\documentclass[fleqn,11pt,twoside]{article}
\usepackage{amsthm,amssymb, color, epsfig, graphics, subfigure}

\usepackage{tikz}

\usepackage{graphicx}
\usepackage{latexsym}
\usepackage[all]{xy}
\usepackage{amsmath, amscd}
\usepackage{graphicx}
\usepackage{lscape}

\newtheorem{proposition}{Proposition}

\makeatletter
\newcommand{\copyrightnote}[2]{{\renewcommand{\thefootnote}{}
 \footnotetext{\small\it
\begin{flushleft}
Copyright \copyright \ #1   #2
\end{flushleft}}}}

\newcommand{\Name}[1]{\begin{flushleft}
                       \LARGE \bf #1
                       \end{flushleft}\vspace{-3mm}}

\newcommand{\Author}[1]{\begin{flushleft}
                       \it #1 \end{flushleft}}

\newcommand{\Address}[1]{\begin{flushleft}
                       \it #1 \end{flushleft}}

\newcommand{\Date}[1]{\begin{flushleft}
                      \small  \it #1 \end{flushleft}}

\newcommand{\evenhead}{Author \ name}
\newcommand{\oddhead}{Article \ name}

\renewcommand{\@evenhead}{
\hspace*{-3pt}\raisebox{-15pt}[\headheight][0pt]{\vbox{\hbox to \textwidth
{\thepage \hfil \evenhead}\vskip4pt \hrule}}}
\renewcommand{\@oddhead}{
\hspace*{-3pt}\raisebox{-15pt}[\headheight][0pt]{\vbox{\hbox to \textwidth
{\oddhead \hfil \thepage}\vskip4pt\hrule}}}
\renewcommand{\@evenfoot}{}
\renewcommand{\@oddfoot}{}

\long\def\@makecaption#1#2{%
  \vskip\abovecaptionskip
  \sbox\@tempboxa{\small \textbf{#1.}\ \ #2}%
  \ifdim \wd\@tempboxa >\hsize
    {\small \textbf{#1.}\ \ #2}\par
  \else
    \global \@minipagefalse
    \hb@xt@\hsize{\hfil\box\@tempboxa\hfil}%
  \fi
  \vskip\belowcaptionskip}

\newcommand{\JNMPnumberwithin}[3][\arabic]{%
  \@ifundefined{c@#2}{\@nocounterr{#2}}{%
    \@ifundefined{c@#3}{\@nocnterr{#3}}{%
      \@addtoreset{#2}{#3}%
      \@xp\xdef\csname the#2\endcsname{%
        \@xp\@nx\csname the#3\endcsname .\@nx#1{#2}}}}%
}

\newcommand{\resetfootnoterule} {
  \renewcommand\footnoterule{%
  \kern-3\p@
  \hrule\@width.4\columnwidth
  \kern2.6\p@}
}

\renewcommand{\footnoterule}{}

\newcommand{\be}{\begin{equation}}
\newcommand{\ee}{\end{equation}}
\newcommand{\ba}{\hspace*{-5pt}\begin{array}}
\newcommand{\ea}{\end{array}}
\newcommand{\p}{\partial}

\makeatother

\numberwithin{equation}{section}
\theoremstyle{definition}

\theoremstyle{proposition}

\renewcommand{\ba}{\begin{array}}
\renewcommand{\ea}{\end{array}}
\newcommand{\beg}{\begin{eqnarray}}
\newcommand{\eeq}{\end{eqnarray}}
\newcommand{\bg}{\begin{eqnarray*}}

\newcommand{\ed}{\end{eqnarray*}}

\newcommand{\nn}{\nonumber}

\renewcommand{\p}{\partial} 
 
\newcommand{\notlhd}{\lhd\kern-.8em{/}\ } 
\newcommand{\notexist}{\ \exists\kern-.5em{\raise.1em\hbox{/}}\ }

\newcommand{\pde}[2]{\frac{\p #1}{\p #2}} 
 
\newcommand{\pdd}[2]{\frac{\p^2 #1}{\p #2^2}} 
\newcommand{\inp}{{\mbox{\vbox{\hrule width0ex\hbox{\vrule
 height0ex\kern3.8pt
\vbox{\kern2.5pt}\kern3.8pt \vrule height1.6ex}
\hrule width1.6ex}}}}

\makeatletter
\newcommand*{\da@rightarrow}{\mathchar"0\hexnumber@\symAMSa 4B }
\newcommand*{\da@leftarrow}{\mathchar"0\hexnumber@\symAMSa 4C }
\newcommand*{\xdashrightarrow}[2][]{%
  \mathrel{%
    \mathpalette{\da@xarrow{#1}{#2}{}\da@rightarrow{\,}{}}{}%
  }%
}
\newcommand{\xdashleftarrow}[2][]{%
  \mathrel{%
    \mathpalette{\da@xarrow{#1}{#2}\da@leftarrow{}{}{\,}}{}%
  }%
}
\newcommand*{\da@xarrow}[7]{%
  \sbox0{$\ifx#7\scriptstyle\scriptscriptstyle\else\scriptstyle\fi#5#1#6\m@th$}%
  \sbox2{$\ifx#7\scriptstyle\scriptscriptstyle\else\scriptstyle\fi#5#2#6\m@th$}%
  \sbox4{$#7\dabar@\m@th$}%
  \dimen@=\wd0 %
  \ifdim\wd2 >\dimen@
    \dimen@=\wd2 %
  \fi
  \count@=2 %
  \def\da@bars{\dabar@\dabar@}%
  \@whiledim\count@\wd4<\dimen@\do{%
    \advance\count@\@ne
    \expandafter\def\expandafter\da@bars\expandafter{%
      \da@bars
      \dabar@ 
    }%
  }%
  \mathrel{#3}%
  \mathrel{%
    \mathop{\da@bars}\limits
    \ifx\\#1\\%
    \else
      _{\copy0}%
    \fi
    \ifx\\#2\\%
    \else
      ^{\copy2}%
    \fi
  }%
  \mathrel{#4}%
}
\makeatother

\begin{document}


\renewcommand{\evenhead}{ {\LARGE\textcolor{blue!10!black!40!green}{{\sf \ \ \ ]ocnmp[}}}\strut\hfill 
M Euler and N Euler}
\renewcommand{\oddhead}{{\LARGE\textcolor{blue!10!black!40!green}{{\sf ]ocnmp[}}}\ \ \ \ \  
From fully-nonlinear to semilinear evolution equations  }

\thispagestyle{empty}
\newcommand{\FistPageHead}[3]{
\begin{flushleft}
\raisebox{8mm}[0pt][0pt]
{\footnotesize \sf
\parbox{150mm}{{\textcolor{blue!10!black!40!green}{{\bf Open Communications in Nonlinear Mathematical Physics}}}
\ \ {Special Issue: Bluman}, 2025\\[0.1cm]
\strut\hfill 
ocnmp:15938,
pp #2\hfill {\sc #3}}}\vspace{-13mm}
\end{flushleft}}

\FistPageHead{1}{\pageref{firstpage}--\pageref{lastpage}}{ \ \ }

\strut\hfill

\strut\hfill

\copyrightnote{The authors. Distributed under Creative Commons Attribution 4.0 International License}

\begin{center}

{\bf {\large A Special OCNMP Issue in Honour of George W Bluman}}
\end{center}

\smallskip


\Name{From fully-nonlinear to semilinear evolution equations: two symmetry-integrable examples\footnote{The Handling Editor at OCNMP was Basil Grammaticos}}

\smallskip

\smallskip

\label{firstpage}

\Author{Marianna Euler and Norbert Euler*}

\Address{
International Society of Nonlinear Mathematical Physics, 
Auf der Hardt 27,\\
56130 Bad Ems, Germany \&
Centro Internacional de Ciencias, Av. Universidad s/n, Colonia Chamilpa,
 62210 Cuernavaca, Morelos, Mexico\\[0.3cm]
* Corresponding author's email address: Dr.Norbert.Euler@gmail.com
}

\Date{Received June 25, 2025; Accepted July 21, 2025}

\smallskip

\smallskip

\noindent
{\bf Abstract:} In this paper we derive two examples of fully-nonlinear symmetry-integrable evolution 
equations with algebraic nonlinearities, namely one class of 3rd-order equations and a 5th-order equation. 
To achieve this we study the equations' Lie-Bäcklund symmetries and apply multipotentialisations, hodograph transformations and generalised hodograph transformations to map the equations to known semilinear integrable evolution equations. As a result of this, we also obtain interesting symmetry-integrable quasilinear equations of order five and order seven, which we display explicitly.

\strut\hfill



\renewcommand{\theequation}{\arabic{section}.\arabic{equation}}

\allowdisplaybreaks

\section{Introduction}

\noindent
In \cite{E-E-76} we reported a set of 3rd-order fully-nonlinear symmetry-intergable equations that are invariant under a projective transformation in $u$  and in \cite{E-E-FN-2022} we reported
some results on 3rd-order fully-nonlinear symmetry-integrable equations
with rational functions in the third derivative. In the current paper we report further results on fully-nonlinear equations whereby we now focus on two special cases, namely a class of 3rd-order evolution equations and a 5th-order evolution equation, both  with algebraic nonlinearities in their highest derivative. In particular, we establish the semi-linearisations of the equations by systematically applying the procedure of multipotentialisation, hodograph transformations and, in some cases, generalised hodograph transformations. We refer to \cite{B-A-Book-2002},  \cite{B-A-C-Book-2010} and \cite{Euler-book-2018} for details on the potentialisation of evolution equations by the use of adjoint symmetries, whereas further details on generalised hodograph tranformations can be found in \cite{E-E-Tree-2001} and  \cite{P-E-E-2004}. 
 
The Schwarzian derivative $S$ plays an important role in our discussion and is defined in terms of the dependent variable $u$ by
 \begin{gather}
 \label{Schwarzian}
 S[u]:=\frac{u_{xxx}}{u_x}-\frac{3}{2}\frac{u_{xx}^2}{u_x^2},
 \end{gather}
 which is applied throughout this paper.
 See for example \cite{Ovsienko} for a discussion on the Schwarzian derivative.
 
The paper is organised as follows: In Section 2 we consider a class of fully-nonlinear 3rd-order equations, identify those equations that admit local Lie-Bäcklund symmetries and map the obtained equations to known semilinear 3rd-order equations. Recursion operators are found for these equations which provide the associated symmetry-integrable hierarchies. In Section 3 we consider a single 5th-order fully-nonlinear evolution equation and establish its Lie-Bäcklund symmetries as well as its mapping to a known 5th-order semilinear integrable equation. In Section 4 we draw our conclusions of the reported results and mention some related open problems.

\section{A class of third-order fully-nonlinear evolution equations}
When seeking symmetry-integrable 3rd-order evolution equations the following statement \cite{E-E-FN-2022} is useful to determine the possible nonlinearities in the highest derivative $u_{xxx}$ \cite{E-E-76}:

\strut\hfill

\noindent
{\bf Lemma 1:} {\it
If a 3rd-order evolution equation of the form
\begin{gather}
u_t=F(x,t,u,u_x,u_{xx},u_{xxx})
\end{gather}
is symmetry-integrable for a given function $F$ then this function must satisfy the following condition:
\begin{gather}
\label{4th-order-cond}
9\left(\pde{F}{u_{xxx}}\right)^2\frac{\p^4F}{\p u_{xxx}^4}
-45\pde{F}{u_{xxx}}\frac{\p^2F}{\p u_{xxx}^2}\frac{\p^3F}{\p u_{xxx}^3}
+40\left(\frac{\p^2F}{\p u_{xxx}^2}\right)^3=0.
\end{gather}
}


\strut\hfill

\noindent
In 
\cite{E-E-FN-2022} we have reported the general solution of (\ref{4th-order-cond}), namely
\begin{gather}
\label{4th-order-cond-GS}
F(x,t,u,u_x,u_{xx},u_{xxx})=\frac{P_3\left(u_{xxx}+P_2\right)}
{\left[(u_{xxx}+P_2)^2+P_1\right]^{1/2}}+P_4,
\end{gather}
where $P_j=P_j(x,t,u,u_x,u_{xx})$, $j=1,2,3,4$, are arbitrary and smooth functions of their arguments. 
We listed several singular solutions of (\ref{4th-order-cond}) that are not included in the general solution 
(\ref{4th-order-cond-GS}) and used the rational solutions in the classification of fully-nonlinear symmetry-integrable equations with rational nonlinearities of order three \cite{E-E-FN-2022}. 

An additional singular solution of 
(\ref{4th-order-cond}) that has not been reported in \cite{E-E-FN-2022} is
\begin{gather}
\label{SS-1}
F(x,t,u,u_x,u_{xx},u_{xxx})=\frac{P_1(x,t,u,u_x,u_{xx})}{\sqrt{u_{xxx}}}+P_2(x,t,u,u_x,u_{xx}),
\end{gather}
which is the form of $F$ that we will focus on in the current paper. Here $P_1$ and $P_2$ are arbitrary and smooth functions.
We furthermore restrict ourselves to evolution equations that do not depend explicitly on their independent variables $x$ and $t$ and apply the standard condition for Lie-Bäcklund symmetries for evolution equations, namely
\begin{gather}
\label{LB-inv}
\left.
\vphantom{\frac{DA}{DB}}
L_E[u]Q\right|_{E=0}=0,
\end{gather}
where $E:=u_t-F(u,u_x,u_{xx},u_{xxx})$, and $Q=Q(u,u_x,u_{xx},\ldots ,u_{nx})$ is the characteristic of the Lie-Bäcklund symmetry generator 
\begin{gather}
\label{IC-LB}
Z_{LB}=Q(u,u_x,u_{xx},\ldots ,u_{nx})\pde{\ }{u},\quad n>3.
\end{gather}
Here $L_E[u]$ denotes the linear operator
\begin{gather}
\label{Lin-Op}
L_E[u]:=\pde{E}{u}+\pde{E}{u_t}D_t+\pde{E}{u_x}D_x+\pde{E}{u_{xx}}D_x^2+\pde{E}{u_{xxx}}D_x^3.
\end{gather}
This leads to 

\strut\hfill

\begin{proposition}
Equation 
\begin{gather}
\label{Eq3-NC}
u_t=
\frac{P_1(u,u_x,u_{xx})}{\sqrt{u_{xxx}}}+P_2(u,u_x,u_{xx})
\end{gather}
is symmetry-integrable if and only if $P_1=\Psi(u_{xx})$ and $P_2=0$, so that 
(\ref{Eq3-NC}) is of the form
\begin{subequations}
\begin{gather}
\label{Main-3rd-EQ}
u_t=\frac{\Psi(u_{xx})}{\sqrt{u_{xxx}}},
\end{gather}
where $\Psi(u_{xx})$ must satisfy the following condition:
\begin{gather}
\label{3rd-Cond}
\Psi\frac{d^2\Psi}{du_{xx}^2}
-\frac{1}{3}\left(\frac{d\Psi}{du_{xx}}\right)^2
+\frac{3\beta}{4}\Psi^{-2/3}=0
\end{gather}
\end{subequations}
with $\beta$ an arbitrary constant.
A hierarchy of symmetry-integrable equations is generated by
\begin{gather}
\label{hier-3rd-order}
u_{t_j}=R^j[u]u_t,\qquad j=1,2,3,\ldots,
\end{gather}
where  $u_t$ is given by (\ref{Main-3rd-EQ}) and 
$R[u]$ is the following recursion operator 
\begin{gather}
\label{R-3rd-order-EQ}
R[u]=G_2D_x^2+G_1D_x+G_0+I_1D_x^{-1}\circ \Lambda_1+I_2D_x^{-1}\circ \Lambda_2
\end{gather}
with
\begin{subequations}
\begin{gather}
G_2=\frac{\Psi^{2/3}}{u_{xxx}}\\[0.3cm]
G_1=\frac{\Psi^{2/3}}{2}\frac{u_{4x}}{u_{xxx}^2}
-\frac{5}{3\Psi^{1/3}}
\frac{d\Psi}{du_{xx}}
\\[0.8cm]
G_0=-\frac{\Psi^{2/3}}{2}
\frac{u_{5x}}{u_{xxx}^2}
+\frac{3\Psi^{2/3}}{4}
\frac{u_{4x}^2}{u_{xxx}^3}
+\frac{1}{3\Psi^{1/3}}
\frac{d\Psi}{du_{xx}}\frac{u_{4x}}{u_{xxx}}
-\frac{8}{9\Psi^{4/3}}
\left(\frac{d\Psi}{du_{xx}}\right)^2
u_{xxx}\nn\\[0.3cm]
\qquad
+\frac{5}{3\Psi^{1/3}}
\frac{d^2\Psi}{du_{xx}^2}
u_{xxx}
+k_0\\[0.3cm]
I_1=\beta\\
\Lambda_1=\frac{u_{4x}}{\Psi^2}
-\frac{2}{\Psi^3}
\frac{d\Psi}{du_{xx}}\\[0.3cm]
I_2=\frac{\Psi}{(u_{xxx})^{1/2}}\\[0.3cm]
\Lambda_2=
\frac{1}{2\Psi^{1/3}}
\frac{u_{6x}}{u_{xxx}^{3/2}}
-\frac{9}{4\Psi^{1/3}}
\frac{u_{4x}u_{5x}}{u_{xxx}^{5/2}}
-\frac{1}{2\Psi^{4/3}}
\frac{d\Psi}{du_{xx}}
\frac{u_{5x}}{u_{xxx}^{1/2}}
-\frac{13}{6\Psi^{4/3}}
\frac{d^2\Psi}{du_{xx}^2}
u_{xxx}^{1/2}u_{4x}\nn\\[0.3cm]
\qquad
+\frac{14}{9 \Psi^{7/3}}
u_{xx}^2u_{xxx}^{1/2}u_{4x}
+\frac{15}{8\Psi^{1/3}}
\frac{u_{4x}^3}{u_{xxx}^{7/2}}
+\frac{5}{12\psi^{4/3}}
\frac{d\Psi}{du_{xx}}
\frac{u_{4x}^2}{u_{xxx}^{3/2}}
+\frac{76}{27\Psi^{7/3}}
\frac{d\Psi}{du_{xx}}
\frac{d^2\Psi}{du_{xx}^2}
u_{xxx}^{5/2}\nn\\[0.3cm]
\qquad
+\frac{41}{27\Psi^{4/3}}
\frac{d^3\Psi}{du_{xx}^3}
u_{xxx}^{5/2}
-\frac{404}{243\Psi^{10/3}}
\left(\frac{d\Psi}{du_{xx}}\right)^3
u_{xxx}^{5/2}
+\frac{2\beta}{\Psi^4}
\frac{d\Psi}{du_{xx}}
u_{xxx}^{5/2}\nn\\[0.3cm]
\qquad
-\frac{\beta}{\Psi^3}u_{xxx}^{1/2}u_{4x}.
\end{gather}
\end{subequations}
Here $k_0$ is an arbitrary constant, and $\Psi(u_{xx})$ and $\beta$ must satisfy condition (\ref{3rd-Cond}).
%
\end{proposition}

\strut\hfill

\noindent
{\bf Proof:} Using the Ansatz (\ref{R-3rd-order-EQ}) with the standard recursion operator condition
(see for example \cite{Euler-book-2018})
\begin{gather}
\left.\vphantom{\frac{DA}{DB}}
\left[L_E[u],\, R[u]\right]=D_tR[u]\right|_{E=0},
\end{gather}
where $E$ defines the equation (\ref{Eq3-NC}) and $L_E$ the linear operator (\ref{Lin-Op}), we find that 
$P_1=\Psi(u_{xx})$, $P_2=0$ with $\Psi$ and $\beta$ that satisfy equation (\ref{3rd-Cond}). This establishes that equation (\ref{Main-3rd-EQ}) is symmetry-integrable under these conditions.
\strut\hfill$\Box$

\smallskip

\noindent
Applying Proposition 1 we find that the second member of the hierarchy (\ref{hier-3rd-order}) is given by the following 5th-order equation:
\begin{gather}
u_{t_1}=-\frac{\Psi^{5/3}}{2}\frac{u_{5x}}{u_{xxx}^{5/2}}
+\frac{5\Psi^{5/3}}{8}\frac{u_{4x}^2}{u_{xxx}^{7/2}}
+\frac{5\Psi^{2/3}}{6}\frac{d\Psi}{du_{xx}}\frac{u_{4x}}{u_{xxx}^{3/2}}
+\frac{4\Psi^{2/3}}{3}
\frac{d^2\Psi}{du_{xx}^2}
u_{xxx}^{1/2}\nn\\[0.3cm]
\qquad
\label{Order-5-Eq-Psi}
-\frac{23}{18\Psi^{1/3}}
\left(\frac{d\Psi}{du_{xx}}\right)^2
u_{xxx}^{1/2}
+\frac{\beta}{\Psi}u_{xxx}^{1/2}
+\frac{k_0\Psi}{u_{xxx}^{1/2}}
\end{gather}

\smallskip

\noindent
For solutions of (\ref{3rd-Cond}) we state the following 

\strut\hfill

\begin{proposition}
Equation (\ref{3rd-Cond}), viz.
\begin{gather*}
\Psi\frac{d^2\Psi}{du_{xx}^2}
-\frac{1}{3}\left(\frac{d\Psi}{du_{xx}}\right)^2
+\frac{3\beta}{4}\Psi^{-2/3}=0,
\end{gather*}
admits the following three solutions:
\begin{enumerate}
\item
For any constant $\beta$ the general solution for (\ref{3rd-Cond})  is
\begin{gather}
\label{Psi-sol-gen}
\Psi(u_{xx})
=(2c_0)^{-3/8}\left[(u_{xx}+c)^2-\beta c_0\right]^{3/4},
\end{gather}
%
%
where $c$ and $c_0$ are arbitrary constants with $c_0\neq 0$. 
\item
For $\beta\neq 0$ an additional solution for (\ref{3rd-Cond}), besides (\ref{Psi-sol-gen}),  is
\begin{gather}
\label{Psi-sol-2}
\Psi(u_{xx})=(2\beta)^{3/8}\left(u_{xx}+c\right)^{3/4},
\end{gather}
where $c$ is an arbitrary constant.
\item
For $\beta=0$ an additional solution for (\ref{3rd-Cond}), besides (\ref{Psi-sol-gen}) with $\beta=0$, is
\begin{gather}
\label{Psi-sol-3}
\Psi(u_{xx})=c,
\end{gather}
where $c$ is an arbitrary constant.
\end{enumerate}

\end{proposition}

\strut\hfill

\noindent
{\bf Remark 1:} {\it 
Equation (\ref{3rd-Cond}) can easily be linearised in a 1st-order ordinary differential equation via a Bernoulli equation. }

\strut\hfill

\noindent
Applying Proposition 1 and Proposition 2 we identify the following three cases of fully-nonlinear 3rd-order symmetry-integrable equations:

\strut\hfill

\noindent
{\bf Case 1.1:} Consider the solution (\ref{Psi-sol-gen}) whereby we let $c_0=1/2$ without loss of generality. This leads to the symmetry-integrable equation 
\begin{gather}
\label{Eq-Case11-1}
u_t=\frac{\displaystyle{\left[(u_{xx}+c)^2-\frac{\beta}{2}\right]^{3/4}}}{\sqrt{u_{xxx}}},
\end{gather}
and by Proposition 1 its recursion operator is given by (\ref{R-3rd-order-EQ}) with
\begin{gather}
\Psi(u_{xx})=\left[(u_{xx}+c)^2-\frac{\beta}{2}\right]^{3/4}
\end{gather}
for any $\beta$ and any constant $c$. 
By introducing the new variable 
$v(x,t)=u_{xx}$, equation (\ref{Eq-Case11-1}) takes the form
\begin{gather}
\label{Eq-Case11-K}
v_t=-\frac{K^{3/4}}{2v_x^{1/2}}
S[v]
-\frac{3v_x^{3/2}}{2K^{5/4}}\left(\frac{v^2}{2}+cv+\frac{c^2}{2}\right)
+\frac{v_x^{3/2}}{K^{1/4}},
\end{gather}
where 
\begin{gather}
K=(v+c)^2-\frac{\beta}{2}.
\end{gather}
We note that, for the case $c=0$ and $\beta=0$, equation (\ref{Eq-Case11-1}) becomes
\begin{gather}
\label{Eq-Case11-1b0}
u_t=\frac{u_{xx}^{3/2}}{\sqrt{u_{xxx}}},
\end{gather}
and equation (\ref{Eq-Case11-K}) becomes
\begin{gather}
v_t=-\frac{1}{2}\left(\frac{v^3}{v_x}\right)^{1/2}S[v]+\frac{3}{4}\left(\frac{v_x^3}{v}\right)^{1/2}.
\end{gather}
In order to establish the semi-linearisation of equation (\ref{Eq-Case11-1}) we need to consider the following two subcases, which distinguishes between the cases where $\beta$ is zero or not:

\strut\hfill

\noindent
{\bf Subcase 1.1a:} Let $\beta=0$. Equation (\ref{Eq-Case11-1}) then takes the following form:
\begin{gather}
\label{Eq-Case11-1-0}
u_t=\frac{(u_{xx}+c)^{3/2}}{\sqrt{u_{xxx}}}.
\end{gather}
By performing a multipotentialisation of (\ref{Eq-Case11-1-0}) we obtain 
\begin{gather}
\label{Eq-Case11-v}
V_t=V^3V_{xxx}-\frac{3}{2^{4/3}}\frac{V_x}{V},
\end{gather}
where 
\begin{gather}
\label{case 11a-V}
V(x,t)=-\frac{1}{2^{1/3}}\left(\frac{u_{xx}+c}{u_{xxx}}\right)^{1/2}.
\end{gather}
We prefer not to show here the details of the multipotentialisation that leads to (\ref{Eq-Case11-v}) but the interested reader can easily verify (\ref{case 11a-V}). 
Applying now the generalised hodograph transformation
\begin{gather}
\label{HT-3}  
{\cal GHT}: 
\left\{
\ba{l}
\displaystyle{
dX=f_1(x,V)  dx+f_2(x,V,V_x,V_{xx})dt  
}\\
\\
\displaystyle{ dT=dt   }\\
\\
\displaystyle{ U(X,T)=x}\\
\ea
\right.
\end{gather}
we obtain the following semilinear equation
\begin{gather}
\label{SKdV-Case11a}
\boxed{\vphantom{\frac{DA}{DB}}
U_T=U_XS[U]+\frac{3}{2^{7/3}} \frac{1}{U_X}
}\ ,
\end{gather}
where 
\begin{subequations}
\begin{gather}
f_1=\frac{1}{V}\\[0.3cm]
f_2=-VV_{xx}+\frac{1}{2}V_x^3
-\frac{3}{2^{7/3}} \frac{1}{V^2}
\end{gather}
\end{subequations}
with $V$ given by (\ref{case 11a-V}).
A 2nd-order recursion operator for equation (\ref{SKdV-Case11a}) has been 
reported in \cite{P-E-E-2004}.

\strut\hfill

\noindent
{\bf Subcase 1.1b:} Let $\beta\neq 0$. Applying the procedure of multipotentialisation on equation 
(\ref{Eq-Case11-1}), {\it viz.}
\begin{gather*}
u_t=\frac{\displaystyle{\left[(u_{xx}+c)^2-\frac{\beta}{2}\right]^{3/4}}}{\sqrt{u_{xxx}}},
\end{gather*}
we obtain
\begin{gather}
\label{case11b-Q-Eq}
Q_t=\frac{1}{Q_x^2}S[Q]-\frac{3}{2^{-7/3}}Q_x^2,
\end{gather}
where
\begin{gather}
\label{case11b-Q-Trans}
Q_x=-\frac{2^{1/3}u_{xxx}}{\displaystyle{\left[(u_{xx}+c)^2+\frac{\beta}{2}\right]^{1/4}}}\ .
\end{gather}
Furthermore equation (\ref{case11b-Q-Eq}) maps to the semilinear equation
\begin{gather}
\label{SKdV-Case11b}
\boxed{\vphantom{\frac{DA}{DB}}
U_T=U_XS[U]-\frac{3}{2^{7/3}}\frac{1}{U_X}
}
\end{gather}
by the standard hodograph transformation
\begin{gather}
\label{HT-case11b}  
{\cal HT}: 
\left\{
\ba{l}
\displaystyle{
X=Q(x,t)
}\\
\\
\displaystyle{ T=t   }\\
\\
\displaystyle{ U(X,T)=x}\\
\ea
\right.
\end{gather}
which completes the semi-linearisation of (\ref{Eq-Case11-1}).
A 2nd-order recursion operator for equation (\ref{SKdV-Case11b}) has been 
reported in \cite{E-E-jnmp-2007}

\strut\hfill

\noindent
{\bf Case 1.2:}  Consider the solution (\ref{Psi-sol-2}) where we let $\beta=1/2$, which is for simplicity but without loss of generality. This leads to the symmetry-integrable equation
\begin{gather}
\label{Eq-Case12-1}
u_t=\frac{\left(u_{xx}+c\right)^{3/4}}{\sqrt{u_{xxx}}},
\end{gather}
whereby its recursion operator is given by (\ref{R-3rd-order-EQ}) with $\beta=1/2$, $c$ an arbitrary constant, and
\begin{gather}
\Psi(u_{xx})=\left(u_{xx}+c\right)^{3/4}.
\end{gather}
By introducing the new variable 
$v(x,t)=u_{xx}$, equation (\ref{Eq-Case12-1}) takes the form
\begin{gather}
v_t=-\frac{1}{2}
\left[\frac{(v+c)^{3/2}}{v_x}\right]^{1/2}S[v]
-\frac{3}{16}\left[\frac{v_x^3}{(v+c)^{5/2}}\right]^{1/2}.
\end{gather}
\noindent
For a semi-linearisation we apply the procedure of multipotentialisation and find that (\ref{Eq-Case12-1}) maps to
\begin{gather}
\label{case12-EqV}
V_t=\frac{S[V]}{V_x^2},
\end{gather}
where 
\begin{gather}
V_x^2=-\frac{u_{xxx}}{2^{2/3}(u_{xx}+c)^{1/2} }\ .
\end{gather}
Furthermore, it is well-known that equation (\ref{case12-EqV}) maps to the Schwarzian KdV
\begin{gather}
\label{SKdV}
\boxed{\vphantom{\frac{DA}{DB}}
U_T=U_XS[U]
}
\end{gather}
by the standard hodograph transformation 
\begin{gather}
\label{HT-case12}  
{\cal HT}: 
\left\{
\ba{l}
\displaystyle{
X=V(x,t)
}\\
\\
\displaystyle{ T=t   }\\
\\
\displaystyle{ U(X,T)=x.}\\
\ea
\right.
\end{gather}
A 2nd-order recursion operator for Schwarzian KdV (\ref{SKdV}) is well known and has for example been 
reported in \cite{E-E-jnmp-2007}

\strut\hfill

\noindent
{\bf Case 1.3:} The symmetry-integrable equation
\begin{gather}
\label{EQ-u-Case13}
u_t=\frac{1}{\sqrt{u_{xxx}}},
\end{gather}
admits the recursion operator (\ref{R-3rd-order-EQ}) with $\Psi=1$ and $\beta=0$. Letting
\begin{gather}
W(x,t)=u_{xxx}
\end{gather}
we obtain
\begin{gather}
\label{EQ-W-Case13}
W_t=\frac{W_{xxx}}{W^{3/2}}
-\frac{9}{2}\frac{W_xW_{xx}}{W^{5/2}}
+\frac{15}{4}\frac{W_x^3}{W^{7/2}}.
\end{gather}
We recall \cite{E-E-1-2025}  that (\ref{EQ-W-Case13}) is also obtained from
\begin{gather}
\label{Eq-u-Til-Case13}
\tilde u_t=-2\frac{\tilde u_x}{\displaystyle{ \sqrt{S[\tilde u]} } }
\end{gather}
where
\begin{gather}
W(x,t)=S[\tilde u]
\end{gather}
so that the relation between $u(x,t)$ and $\tilde u(x,t)$ is
\begin{gather}
u_{xxx}=S[\tilde u].
\end{gather}
In \cite{E-E-1-2025} we established that (\ref{Eq-u-Til-Case13}) maps to the Schwarzian KdV with a hodograph-type transformation. Using this result, we find that equation (\ref{EQ-u-Case13}) maps to the Schwarzian KdV (\ref{SKdV}), {\it viz.}
\begin{gather*}
\boxed{\vphantom{\frac{DA}{DB}}
U_T=U_XS[U]
}
\end{gather*}
under the following change of variables:
\begin{gather}
\label{HT-2}  
{\cal HT}: 
\left\{
\ba{l}
\displaystyle{
X=\int \sqrt{u_{xxx}}\  dx  
}\\
\\
\displaystyle{ T=t   }\\
\\
\displaystyle{ U(X,T)=x.}\\
\ea
\right.
\end{gather}

\noindent
For the sake of completeness, we give the recursion operator for the equation
\begin{gather}
U_T=U_XS[U]+\frac{\lambda}{U_X},
\end{gather}
which includes the semilinear equations (\ref{SKdV-Case11a}) for $\lambda=3\cdot 2^{-7/3}$, (\ref{SKdV-Case11b}) for $\lambda=-2\cdot 3^{-7/3}$, and (\ref{SKdV}) for $\lambda=0$, namely \cite{E-E-jnmp-2007}
\begin{gather}
R[U]=D_X^2-\frac{2U_{XX}}{U_X}D_X
+\frac{U_{XXX}}{U_X}
-\frac{U_{XX}^2}{U_X^2}
-\frac{2\lambda}{3U_X^2}+k_0
-\frac{8\lambda}{3}D_X^{-1}\circ \frac{U_{XX}}{U_X^3}\nn\\[0.3cm]
\qquad
-U_XD_X^{-1}\circ\left(
\frac{U_{4X}}{U_X^2}
-\frac{4U_{XX}U_{XXX}}{U_X^3}
+\frac{3U_{XX}^3}{U_X^4}
-\frac{2\lambda U_{XX}}{U_X^4}\right).
\end{gather}

\section{A fifth-order fully-nonlinear evolution equation}
When seeking symmetry-integrable 5th-order evolution equations, the following statement \cite{E-E-FN-2022} is useful to determine the possible nonlinearities in the highest derivative $u_{5x}$ \cite{E-E-1-2025}:

\strut\hfill

\noindent
{\bf Lemma 2:} {\it 
If a 5th-order evolution equation of the form
\begin{gather}
\label{Cond-5th-order-eq}
u_t=F(x,t,u,u_x,\ldots,u_{5x})
\end{gather}
is symmetry-integrable for a given function $F$ then this function must satisfy the following condition:
\begin{gather}
\label{3rd-order-cond}
5\pde{F}{u_{5x}}\frac{\p^3 F}{\p u_{5x}^3}-8\left(\pdd{F}{u_{5x}}\right)^2=0.
\end{gather}
}

\strut\hfill

\noindent
The general solution of (\ref{3rd-order-cond}) is 
\begin{gather}
F(x,t,u,u_x,\ldots,u_{5x})=\frac{F_1}{(u_{5x}+F_2)^{2/3}}+F_3,
\end{gather}
where $F_j=F_j(x,t,u,u_x,\ldots,u_{4x})$ $(j=1,2,3)$ are arbitrary and smooth functions of their arguments. In the current paper we consider the special case where $F_1=1,$ $F_2=F_3=0$. Applying  now the Lie-Bäcklund symmetry invariance condition (\ref{LB-inv}) leads to the following

\strut\hfill

\begin{proposition}
The fully-nonlinear 5th-order equation
\begin{gather}
\label{Main-5th-EQ-u}
u_t=\frac{1}{u_{5x}^{2/3}}
\end{gather}
is symmetry-integrable and admits Lie-Bäcklund symmetries of order $1+6n$ and order $5+6n$, for every natural number $n$, so the symmetries are of order $\{7,11,13,17,19, \ldots\}$. 

\end{proposition}

\strut\hfill

\noindent
Let us give the first two equations in this hierarchy explicitly:
The symmetry-integrable evolution equation associated with the 7th-order Lie-Bäcklund symmetry of equation
 (\ref{Main-5th-EQ-u}) is 
\begin{gather}
\label{7th-Eq-u}
u_{t}=\frac{u_{7x}}{u_{5x}^{7/3}}-\frac{7}{6}\frac{u_{6x}^2}{u_{5x}^{10/3}}
\end{gather}
and the associated 11th-order symmetry-integrable equation is
\begin{gather}
u_t=\frac{1}{u_{5x}^{11/3}}\left(u_{11x}
-\frac{11u_{6x}u_{10x}}{u_{5x}}
-\frac{77 u_{7x}u_{9x}}{3u_{5x}}
+\frac{682 u_{6x}^2u_{9x}}{9u_{5x}^{2}}
-\frac{33 u_{8x}^2}{2u_{5x}}
-\frac{374 u_{6x}^3u_{8x}}{u_{5x}^{3}}\right.\nn\\[0.3cm]
\qquad
\left.
+\frac{286u_{6x}u_{7x}u_{8x}}{u_{5x}^2}
+\frac{1892 u_{7x}^3}{27u_{5x}^{2}}
-\frac{22066 u_{6x}^2u_{7x}^2}{27 u_{5x}^{3}}
+\frac{107525 u_{6x}^4u_{7x}}{81u_{5x}^{4}}
-\frac{752675 u_{6x}^6}{1458u_{5x}^{5}}\right).
\end{gather}

\noindent
We now turn to the task to semi-linearise equation (\ref{Main-5th-EQ-u}): We apply the multipotentialisation procedure systematically on equation (\ref{Main-5th-EQ-u}), which leads to the quasilinear equation
\begin{gather}
\label{5th-EQ-w}
w_t=\frac{w_{5x}}{w_x^5}-\frac{10w_{xx}w_{4x}}{w_x^6}
-\frac{10w_{xxx}^2}{w_x^6}
+\frac{60w_{xx}^2w_{xxx}}{w_x^7}
-\frac{45w_{xx}^4}{w_x^8},
\end{gather}
where
\begin{gather}
w_x^3=-\frac{3}{2}u_{5x}.
\end{gather}
Equation (\ref{5th-EQ-w}) then takes the quasilinear form
\begin{gather}
\label{5th-EQ-V}
V_t=V^5V_{5x}+5V^4V_xV_{4x}+10V^4V_{xx}V_{xxx},
\end{gather}
with the following change of variables:
\begin{gather}
V(x,t)=\frac{1}{w_x}.
\end{gather}
Using generalised hodograph transformations we conclude that
%
%
the fully-nonlinear 5th-order equation (\ref{Main-5th-EQ-u})
maps to the semilinear 5th-order equation
\begin{gather}
\label{3rd-Pot-Kuper}
\boxed{\vphantom{\frac{DA}{DB}}
U_T=U_X\left(\frac{\p^2 S[U]}{\p X^2}+4S^2[U]\right)
}
\end{gather}
by the generalised hodograph transformation
\begin{gather}
{\cal GHT}: 
\left\{
\ba{l}
\displaystyle{
dX=f_1(x,V)  dx+f_2(x,V,V_x,V_{xx},V_{xxx},V_{4x})dt  
}\\
\\
\displaystyle{ dT=dt   }\\
\\
\displaystyle{ U(X,T)=x,}\\
\ea
\right.
\end{gather}
where
\begin{subequations}
\begin{gather}
f_1=\frac{1}{V}\\[0.3cm]
f_2=-V^3V_{4x}-2V^2V_xV_{xxx}-4V^2V_{xx}^2+4VV_{x}^2V_{xx}-V_x^4
\end{gather}
and 
\begin{gather}
V(x,t)=-\left(\frac{2}{3}\right)^{1/5}\frac{1}{u_{5x}^{1/3}}.
\end{gather}
\end{subequations}


\strut\hfill

\noindent
Some remarks are in order:

\strut\hfill

\noindent
{\bf Remark 2:} {\it
We find  that equation (\ref{Main-5th-EQ-u}) does not admit a recursion operator of the form
\begin{gather}
\label{6th-R-ansatz}
R[u]=\sum_{j=0}^6 G_jD_x^j+\sum_{j=1}^3 I_jD_x^{-1}\circ \Lambda_j,
\end{gather}
whereby we assumed the following dependencies: for the integrating factors $\Lambda_k=\Lambda_k(u,u_x,$ $u_{xx},\ldots,u_{12x})$; for the symmetry coefficients $I_j=I_j(u,u_x,u_{xx},\ldots,u_{5x})$, and for the coefficients of $D_x^j$ we assumed that there are no restrictions in the number of derivatives, i.e. $G_j=G_j(u,u_x,$ $u_{xx},\ldots)$. Since a recursion operator of this form does not exist, we expect that the recursion operator is of a nonlocal type.
We will not explore this further here.}

\strut\hfill

\noindent
{\bf Remark 3:}
{\it The semilinear equation (\ref{3rd-Pot-Kuper}) is the third potentialisation  (see equation (2.6) \cite{Euler-book-2018})  in the chain of multipotentialisations of the Kupershmidt equation 
\begin{gather}
K_t=K_{5x}-5K_xK_{xxx}-5K_{xx}^2-5K^2K_{xxx}
-20KK_xK_{xx}
-5K_x^3+5K^4K_x.
\end{gather}
A recursion operator of order six has been reported for equation (\ref{3rd-Pot-Kuper}) in \cite{Euler-book-2018}, where some solution-generating formulas are also given for (\ref{3rd-Pot-Kuper}).

}

\strut\hfill

\noindent
{\bf Remark 4:}
{\it Let us furthermore point out that equation (\ref{5th-EQ-V}), viz.
\begin{gather*}
V_t=V^5V_{5x}+5V^4V_xV_{4x}+10V^4V_{xx}V_{xxx},
\end{gather*}
just like equation (\ref{Main-5th-EQ-u}), does not admit a recursion operator of the form (\ref{6th-R-ansatz}) although (\ref{5th-EQ-V}) is of course symmetry-integrable and hence admits Lie-Bäcklund symmetries of the same order as equation (\ref{5th-EQ-w}) which generates a hierarchy of symmetry-integrable equations. For example, the 7th-order symmetry-integrable equation in this hierarchy is
\begin{gather}
V_t=V^7V_{7x}
+14V^6V_xV_{6x}
+49V^5V_x^2V_{5x}
+28V^6V_{xx}V_{5x}
+35V^4V_x^3V_{4x}
+140V^5V_xV_{xx}V_{4x}\nn\\[0.3cm]
\label{7th-Eq-V}
\qquad
+35V^6V_{xxx}V_{4x}
+70V^5V_xV_{xxx}^2
+70V^5V_{xx}^2V_{xxx}
+70V^4V_x^2V_{xx}V_{xxx}.
\end{gather}
Furthermore, the 7th-order equation (\ref{7th-Eq-u}) maps to the 7th-order equation (\ref{7th-Eq-V}) by the following change of variable:
\begin{gather}
V(x,t)=u_{5x}^{-1/3}.
\end{gather}
}

\noindent
{\bf Remark 5:}
{\it Quite remarkably, equation (\ref{5th-EQ-w}), viz.
\begin{gather*}
w_t=\frac{w_{5x}}{w_x^5}-\frac{10w_{xx}w_{4x}}{w_x^6}
-\frac{10w_{xxx}^2}{w_x^6}
+\frac{60w_{xx}^2w_{xxx}}{w_x^7}
-\frac{45w_{xx}^4}{w_x^8},
\end{gather*}
admits the following 6th-order recursion operator:
\begin{gather}
\label{6th-R-ansatz}
R[w]=\sum_{j=0}^6 G_jD_x^j+2w_tD_x^{-1}\circ \Lambda_1+2D_x^{-1}\circ \Lambda_2,
\end{gather}
where
\begin{subequations}
\begin{gather}
G_6=\frac{1}{w_x^6}\\[0.3cm]
G_5=-\frac{15w_{xx}}{w_x^7}\\[0.3cm]
G_4=-\frac{32w_{xxx}}{w_x^7}+\frac{123 w_{xx}^2}{w_x^8}\\[0.3cm]
G_3=-\frac{27w_{4x}}{w_x^7}
+\frac{354 w_{xx}w_{xxx}}{w_x^8}
-\frac{600w_{xx}^3}{w_x^9}\\[0.3cm]
G_2=-\frac{19w_{5x}}{w_x^7}
+\frac{271w_{xx}w_{4x}}{w_x^8}
+\frac{232w_{xxx}^2}{w_x^8}
-\frac{2040w_{xx}^2w_{xxx}}{w_x^9}
+\frac{1980w_{xx}^4}{w_x^{10}}\\[0.3cm]
G_1=-\frac{4w_{6x}}{w_x^7}
+\frac{79w_{xx}w_{5x}}{w_x^8}
+\frac{149w_{xxx}w_{4x}}{w_x^8}
-\frac{754w_{xx}^2w_{4x}}{w_x^9}
+\frac{4440w_{xx}^3w_{xxx}}{w_x^{10}}\nn\\[0.3cm]
\qquad
+\frac{1126w_{xx}w_{xxx}^2}{w_x^9}
-\frac{3060w_{xx}^5}{w_x^{11}}\\[0.3cm]
G_0=-\frac{3w_{7x}}{w_x^7}
+\frac{63w_{xx}w_{6x}}{w_x^8}
-\frac{690w_{xx}^2w_{5x}}{w_x^9}
+\frac{145 w_{xxx}w_{5x}}{w_x^8}
-\frac{2494 w_{xx}w_{xxx}w_{4x}}{w_x^9}\nn\\[0.3cm]
\qquad
+\frac{4794 w_{xx}^3w_{4x}}{w_x^{10}}
+\frac{87w_{4x}^2}{w_x^8}
-\frac{632w_{xxx}^3}{w_x^9}
-\frac{22680 w_{xx}^4w_{xxx}}{w_x^{11}}
+\frac{10326 w_{xx}^2w_{xxx}^2}{w_x^{10}}\nn\\[0.3cm]
\qquad
+\frac{11340 w_{xx}^6}{w_x^{12}}+k_0\\[0.3cm]
\Lambda_1=-\frac{w_{4x}}{w_x^3}+\frac{6w_{xx}w_{xxx}}{w_x^4}-\frac{6w_{xx}^3}{w_x^5}\\[0.3cm]
\Lambda_2=\frac{w_{8x}}{w_x^7}
-\frac{28w_{xx}w_{7x}}{w_x^8}
+\frac{396w_{xx}^2w_{6x}}{w_x^9}
-\frac{68w_{xxx}w_{6x}}{w_x^8}
-\frac{106u_{4x}u_{5x}}{w_x^8}
-\frac{3636w_{xx}^3w_{5x}}{w_x^{10}}\nn\\[0.3cm]
\qquad
+\frac{1656w_{xx}w_{xxx}w_{5x}}{w_x^9}
+\frac{1060w_{xx}w_{4x}^2}{w_x^9}
+\frac{23265w_{xx}^4w_{4x}}{w_x^{11}}
+\frac{1420w_{xxx}^2w_{4x}}{w_x^9}\nn\\[0.3cm]
\qquad
-\frac{18900 w_{xx}^2w_{xxx}w_{4x}}{w_x^{10}}
-\frac{8520w_{xx}w_{xxx}^3}{w_x^{10}}
+\frac{63360w_{xx}^3w_{xxx}^2}{w_x^{11}}
-\frac{103950 w_{xx}^5w_{xxx}}{w_x^{12}}\nn\\[0.3cm]
\qquad
+\frac{44550w_{xx}^7}{w_x^{13}}.
\end{gather}
Here $k_0$ is an arbitrary constant.
\end{subequations}
It is interesting to note that acting $R[w]$ on $w_x$ does not result in a 7th-order symmetry for
(\ref{5th-EQ-w}). Instead, $R[w]$ maps the $x$-translation symmetry to itself, i.e.,
\begin{gather}
R[w]\,w_x=k_0w_x.
\end{gather}
Nevertheless, we find that equation (\ref{5th-EQ-w}) does admit a 7th-order Lie-Bäcklund symmetry and hence a corresponding 7th-order symmetry-integrable equation, namely
\begin{gather}
w_t=\frac{w_{7x}}{w_x^8}
-\frac{21w_{xx}w_{6x}}{w_x^8}
-\frac{49w_{xxx}w_{5x}}{w_x^8}
+\frac{231w_{xx}^2w_{5x}}{w_x^9}
-\frac{28w_{4x}^2}{w_x^8}
+\frac{826w_{xx}w_{xxx}w_{4x}}{w_x^9}\nn\\[0.3cm]
\label{7th-Eq-w2}
\qquad
-\frac{1596w_{xx}^3w_{4x}}{w_x^{10}}
+\frac{644w_{xxx}^3}{3w_x^9}
-\frac{3444w_{xx}^2w_{xxx}^2}{w_x^{10}}
+\frac{7560 w_{xx}^4w_{xxx}}{w_x^{11}}
-\frac{3780 w_{xx}^6}{w_x^{12}}.
\end{gather}
With the change of variable 
\begin{gather}
W(x,t)=\frac{1}{w_x}
\end{gather}
equation (\ref{7th-Eq-w2}) takes the form
\begin{gather}
W_t=W^8W_{7x}
-6W^7W_xW_{6x}
+21W^6W_xW_{6x}
-42W^7W_{xx}W_{5x}
+70W^6W_{xx}W_{5x}\nn\\[0.3cm]
\qquad
+7W^5W_x^2W_{5x}
+30W^6W_x^2W_{5x}
-70W^7W_{xxx}W_{4x}
+105W^6W_{xxx}W_{4x}\nn\\[0.3cm]
\qquad
+390W^6W_xW_{xx}W_{4x}
-280W^5W_xW_{xx}W_{4x}
-120W^5W_x^3W_{4x}
+245W^4W_x^3W_{4x}\nn\\[0.3cm]
\qquad
+260W^6W_xW_{xxx}^2
-210W^5W_xW_{xxx}^2
+630W^6W_{xx}^2W_{xxx}
-540W^5W_{xx}^2W_{xxx}\nn\\[0.3cm]
\qquad
-2160W^5W_x^2W_{xx}W_{xxx}
+2510W^4W_x^2W_{xx}W_{xxx}
+360W^4W_x^4W_{xxx}\nn\\[0.3cm]
\qquad
-760W^3W_x^4W_{xxx}
-1800W^5W_xW_{xx}^3
+1830W^4W_xW_{xx}^3
+3960W^4W_x^3W_{xx}^2\nn\\[0.3cm]
\qquad
\label{7th-Eq-W}
-4800W^3W_x^3W_{xx}^2
-720W^3W_x^5W_{xx}
+2280W^2W_x^5W_{xx}
-720W^2W_x^7.
\end{gather}
}

\section{Concluding remarks}
In this paper we have introduced the following set of fully-nonlinear evolution equations:
\begin{gather}
\label{FN-Set}
u_t=\frac{\displaystyle{\left[(u_{xx}+c)^2-\frac{\beta}{2}\right]^{3/4}}}{\sqrt{u_{xxx}}},\quad
u_t=\frac{\displaystyle{\left(u_{xx}+c\right)^{3/4}}}{\sqrt{u_{xxx}}},\quad
u_t=\frac{1}{\sqrt{u_{xxx}}},\quad
u_t=\frac{1}{u_{5x}^{2/3}}.
\end{gather}
We have established that this set of equations is symmetry-integrable in the sense that the equations admit local Lie-Bäcklund symmetries and that the equations can be mapped to known semilinear integrable equations using a combination of multipotentialisations, hodograph transformations and generalised hodegraph transformations. The 3rd-order fully-nonlinear equations in the set (\ref{FN-Set}) admit recursion operators of the standard 2nd-order form which can be applied to generated hierarchies of symmetry-integrable quasilinear equations. On the other hand, according to our calculations, we find that the fully-nonlinear 5th-order equation listed in (\ref{FN-Set}) does not admit a standard recursion operator of order six or less. It is therefore an open problem to find a recursion operator for this fully-nonlinear 5th-order equation, which we expect to be nonlocal.

Besides the fully-nonlinear equations listed in (\ref{FN-Set}) we have obtained a set of quasilinear 5th-order symmetry-integrable equations, namely the equation (\ref{Order-5-Eq-Psi}) under the condition (\ref{3rd-Cond}), as well as the equations (\ref{5th-EQ-w}) and (\ref{5th-EQ-V}). Furthermore, we have obtained
a set of 7th-order quasilinear symmetry-integrable equations, namely (\ref{7th-Eq-u}), (\ref{7th-Eq-V}), (\ref{7th-Eq-w2}) and (\ref{7th-Eq-W}). These equations result naturally from the multipotentialisations of the fully-nonlinear equations (\ref{FN-Set}) and as members of the symmetry-integrable hierarchies. 


Finally we should mention that the complete classification of fully-nonlinear symmetry-integrable evolution equations of 3rd and higher order is an ongoing project and the examples 
studied here is an addition to the results that have already been reported in \cite{E-E-76}, 
\cite{E-E-FN-2022} and 
\cite{E-E-1-2025}. 

\label{lastpage}

\end{document}